\documentclass[prd,superscriptaddress,showpacs,nofootinbib,amsmath,amssymb,aps,10pt]{revtex4}

\usepackage{bm}
\usepackage{amsfonts}
\usepackage{latexsym}
\usepackage[latin1]{inputenc}
\usepackage{graphicx}
\usepackage{amsmath}
\usepackage{palatino}
\usepackage{mathpazo}
\usepackage{booktabs}
\usepackage{dcolumn}

\usepackage{wrapfig}
\usepackage{wasysym}
\usepackage[outdir=./]{epstopdf}

\def\jnl@style{\it}
\def\aaref@jnl#1{{\jnl@style#1}}

\def\aaref@jnl#1{{\jnl@style#1}}

\def\aj{\aaref@jnl{AJ}}                   % Astronomical Journal
\def\apj{\aaref@jnl{ApJ}}                 % Astrophysical Journal
\def\apjl{\aaref@jnl{ApJ}}                % Astrophysical Journal, Letters
\def\apjs{\aaref@jnl{ApJS}}               % Astrophysical Journal, Supplement
\def\apss{\aaref@jnl{Ap\&SS}}             % Astrophysics and Space Science
\def\aap{\aaref@jnl{A\&A}}                % Astronomy and Astrophysics
\def\aapr{\aaref@jnl{A\&A~Rev.}}          % Astronomy and Astrophysics Reviews
\def\aaps{\aaref@jnl{A\&AS}}              % Astronomy and Astrophysics, Supplement
\def\mnras{\aaref@jnl{Mon.~Not.~Roy.~Astron.~Soc.}}             % Monthly Notices of the RAS
\def\prd{\aaref@jnl{Phys.~Rev.~D}}        % Physical Review D
\def\prc{\aaref@jnl{Phys.~Rev.~C}}  % Physical Review C
\def\prl{\aaref@jnl{Phys.~Rev.~Lett.}}    % Physical Review Letters
\def\qjras{\aaref@jnl{QJRAS}}             % Quarterly Journal of the RAS
\def\skytel{\aaref@jnl{S\&T}}             % Sky and Telescope
\def\ssr{\aaref@jnl{Space~Sci.~Rev.}}     % Space Science Reviews
\def\zap{\aaref@jnl{ZAp}}                 % Zeitschrift fuer Astrophysik
\def\nat{\aaref@jnl{Nature}}              % Nature
\def\aplett{\aaref@jnl{Astrophys.~Lett.}} % Astrophysics Letters
\def\apspr{\aaref@jnl{Astrophys.~Space~Phys.~Res.}} % Astrophysics Space Physics Research
\def\physrep{\aaref@jnl{Phys.~Rep.}}      % Physics Reports
\def\physscr{\aaref@jnl{Phys.~Scr}}       % Physica Scripta
\def\commat{\aaref@jnl{Comm.~Math.~Phys.}}              % Communications in Mathematical Physics
\def\science{\aaref@jnl{Science}}               % Science
\def\cqg{\aaref@jnl{Classical Quant.~Grav.}}            % Classical and Quantum Gravity
\def\jpcs{\aaref@jnl{JPCS}}                                     % Journal of Physics Conference Series
\def\ijmpd{\aaref@jnl{Int.~J.~Mod.~Phys.~D}}                    % International Journal of Modern Physics D
\def\grg{\aaref@jnl{Gen.~Relat.~Gravit.}}               % General Relativity and Gravitation
\def\rpp{\aaref@jnl{Rep.~Prog.~Phys.}}          % Reports on Progress in Physics
\def\npa{\aaref@jnl{Nucl.~Phys.~A}}        % Nuclear Physics A
\def\lrr{\aaref@jnl{Living Rev.~Rel.}}                   % Living reviews in relativity
\def\jcap{\aaref@jnl{J.~Cosmology Astropart.~Phys.}}    % Journal of cosmology and astroparticle physics
\def\rmp{\aaref@jnl{Rev.~Mod.~Phys.}}   %Reviews of modern physics

\allowdisplaybreaks[1]
\begin{document}

\title{Slowly rotating neutron and strange stars in $R^2$  gravity}

\author{Kalin V. Staykov}
\email{kalin.v.staikov@gmail.com}
\affiliation{Department of Theoretical Physics, Faculty of Physics, Sofia University, Sofia 1164, Bulgaria}
\affiliation{Theoretical Astrophysics, Eberhard Karls University of T\"ubingen, T\"ubingen 72076, Germany}

\author{Daniela D. Doneva}
\email{daniela.doneva@uni-tuebingen.de}
\affiliation{Theoretical Astrophysics, Eberhard Karls University of T\"ubingen, T\"ubingen 72076, Germany}
\affiliation{INRNE - Bulgarian Academy of Sciences, 1784  Sofia, Bulgaria}

\author{Stoytcho S. Yazadjiev}
\email{yazad@phys.uni-sofia.bg}
\affiliation{Department of Theoretical Physics, Faculty of Physics, Sofia University, Sofia 1164, Bulgaria}
\affiliation{Theoretical Astrophysics, Eberhard Karls University of T\"ubingen, T\"ubingen 72076, Germany}

\author{Kostas D. Kokkotas}
\email{kostas.kokkotas@uni-tuebingen.de}
\affiliation{Theoretical Astrophysics, Eberhard Karls University of T\"ubingen, T\"ubingen 72076, Germany}
\affiliation{Department of Physics, Aristotle University of Thessaloniki, Thessaloniki 54124, Greece}

%%%%%%%%%%%%%%%%%%%%%%%%%%%%%%%%%%%%  DATE  %%%%%%%%%%%%%%%%%%%%%%%%%%%%%%%%%%%%

\begin{abstract}
In the present paper we investigate self-consistently  slowly rotating neutron and strange stars in R-squared gravity. For this purpose we first derive the equations describing the
structure of the slowly rotating compact stars in $f(R)$-gravity and then simultaneously solve the exterior and the interior problem. The structure of the slowly rotating
neutron stars is studied for  two different hadronic equations of state and a strange matter equation of state. The moment of inertia and its dependence on the stellar mass
and the $R$-squared gravity parameter $a$ is also examined in details.  We find that the neutron star moment of inertia for large values of the parameter $a$ can be up to $30\%$  larger compared to the corresponding general relativistic models. This is much higher than the change in the maximum mass induced by $R$-squared gravity and is beyond the EOS uncertainty. In this way the future observations of the moment of inertia of compact stars  could allow us
to distinguish between  general relativity and $f(R)$  gravity, and more generally to test the strong field regime of gravity.

\end{abstract}

\maketitle
\date{}
\section{Introduction}

The $f(R)$ theories are natural generalizations of Einstein's theory and they are widely explored alternative theories of gravity trying to explain the accelerated expansion of the universe.
Their essence is that the standard Einstein-Hilbert Lagrangian is replaced by a  function of the Ricci scalar curvature $R$. Many different classes of $f(R)$ theories were constructed and examined (for a review see \cite{Sotiriou2010,DeFelice2010a,Nojiri2011}). In the current paper we will be concentrated on the so-called $R^2$-gravity,  where  the standard Einstein-Hilbert Lagrangian is replaced by  $R + a R^2$.

Even though the $f(R)$ theories are normally employed to explain cosmological observation, the various astrophysical phenomena can also be used to impose constraints on these theories. Some of the most suitable objects in this direction are the neutron stars because of their high compactness and the rich spectrum of observations. Moreover the neutron stars can serve  as a tool to test the strong field regime of the alternative theories of gravity. As the  investigations show,  nonlinear effects can appear in alternative theories of gravity when strong fields are considered for both neutron stars and black holes \cite{Damour1993,Stefanov2008,Doneva2010,Doneva2013}, which are not present for weak fields.

Static neutron stars in $f(R)$ theories of gravity were examined up to now in many papers \cite{Cooney2010,Babichev2010,Arapoglu2011,Jaime2011,Santos2012,Orellana2013,Alavirad2013,Astashenok2013}. However, most of these studies were based on a perturbative and not self-consistent
method. The non-perturbative  and self-consistent study in \cite{Yazadjiev2014} shows that the perturbative approach in the parameter $a$ is not suitable and one should solve the field equations self-consistently in order to obtain reliable physical results. From the results in \cite{Yazadjiev2014} one can conclude that the mass and the radii of neutron stars can increase considerably  for certain values of the parameters, but these changes are still comparable to the changes induced by different nuclear equations of state (EOSs). So the observations of the neutron star masses and radii alone can not be used  to put constraints on the free parameters of the theory.

Our purpose in the current paper is to take the first steps toward thoroughly studying rotating compact stars  in $f(R)$ theories of gravity and their
astrophysical manifestations. Similar studies were already performed in other generalized theories of gravity (see e.g. \cite{Damour1996,Sotani2010,Sotani2012,Doneva2013}). A very first step in this direction and an important extension of the results in \cite{Yazadjiev2014} is to consider the slow rotation approximation (in linear order of the angular velocity $\Omega$). Even though this approximation is expected to be valid only for rotational frequencies below a few hundred Hz, it turns out that most of the observed neutron stars fall into this category. A drawback is that it can not account for the changes in the mass and radius of the neutron stars due to rotation, because these effects are of the order $\Omega^2$. But the slow rotation approximation can give us information about the frame dragging  around neutron stars and the moment of inertial. This can be already observationally relevant, because it is expected that in the near future the observations of double neuron stars would allow us to measure the moment of inertia with a good accuracy \cite{Lattimer2005,Kramer2009}. Moreover the slow rotation approximation can be used to study the $r$-modes for rotating neutron stars and the associated gravitational wave emission \cite{Andersson98b,Friedman98,Kokkotas99b}.

In the present paper we study models with both nuclear matter EOS and strange star EOS. Up to our knowledge this is the first self-consistent study of strange stars in $f(R)$ theories of gravity. It is important because strange stars have very distinct properties compared to the standard neutron stars and our investigations can help us to  determine up to what extend the results and conclusions in the present paper are EOS independent.

Studying rapidly rotating neutron stars is also important, because as the results in the case of scalar-tensor theories of gravity \cite{Doneva2013} show, rapid rotation can magnify significantly the deviations from general relativity. It is expected that the same will be true also for rapidly rotating neutron stars in $f(R)$ theories, but this  study is very complex and involved and we leave it for a future publication.

\section{Basic equations}

The $f(R)$ theories are described by the following action
\begin{eqnarray}\label{A}
S= \frac{1}{16\pi G} \int d^4x \sqrt{-g} f(R) + S_{\rm
matter}(g_{\mu\nu}, \chi),
\end{eqnarray}
with  $R$ being the scalar curvature with respect to the spacetime
metric $g_{\mu\nu}$. $S_{\rm matter}$ is the action of the matter
fields collectively denoted by $\chi$. In order for the $f(R)$ theories to be free
of tachyonic instabilities and the appearance of ghosts, the following inequalities have to be satisfied  \cite{Sotiriou2010,DeFelice2010a,Nojiri2011}
\begin{eqnarray}
\frac{d^2f}{dR^2}\ge 0,  \;\;\; \frac{df}{dR}>0,
\end{eqnarray}
respectively. In the spacial case of $R^2$ gravity the above inequalities give $a\ge 0$ and $1+ 2aR\ge 0$.

It is well-known that the $f(R)$ theories are mathematically equivalent to the Brans-Dicke theory (with $\omega_{BD}=0$) given by the action
\begin{eqnarray}
S=\frac{1}{16\pi G} \int d^4x \sqrt{-g}\left[\Phi R - U(\Phi)\right]
+ S_{\rm matter}(g_{\mu\nu}, \chi),
\end{eqnarray}
where the gravitational scalar $\Phi$  and the  potential $U(\Phi)$ are defined by $\Phi=\frac{df(R)}{dR}$ and
$U(\Phi)=R \frac{df}{dR} - f(R)$, respectively.  In the case of $R^2$ gravity we have $\Phi=1 + 2aR$ and  the Brans-Dicke potential is $U(\Phi)=\frac{1}{4a}(\Phi-1)^2$.

In many cases it proved useful to study the scalar-tensor theories in the so-called Einstein frame with metric  $g^{*}_{\mu\nu}$ defined by the conformal transformation
$g^{*}_{\mu\nu}=\Phi g_{\mu\nu}$. The Einstein frame action can be written in the form

\begin{eqnarray}\label{EFA}
S=\frac{1}{16\pi G} \int d^4x \sqrt{-g^{*}}\left[ R^{*} - 2
g^{*\mu\nu}\partial_{\mu}\varphi \partial_{\nu}\varphi - V(\varphi)
\right] + S_{\rm
matter}(e^{-\frac{2}{\sqrt{3}}\varphi}g^{*}_{\mu\nu},\chi),
\end{eqnarray}
where $R^{*}$ is the Ricci scalar curvature with respect to the
Einstein frame metric $g^{*}_{\mu\nu}$ and the new scalar field $\varphi$ is defined by $\varphi =\frac{\sqrt{3}}{2}\ln\Phi$. The Einstein frame potential $V(\varphi)$ is correspondingly
$V(\varphi)=A^4(\varphi)U(\Phi(\varphi))$  with $A(\varphi)$ defined as $A^2(\varphi)=\Phi^{-1}(\varphi)=
e^{-\frac{2}{\sqrt{3}}\varphi}$.
 In the case of $R^2$ gravity one can also show that $V(\varphi)= \frac{1}{4a}
\left(1-e^{-\frac{2\varphi}{\sqrt{3}}}\right)^2$.

In accordance with the main purpose of the present
paper we consider stationary and axisymmetric spacetimes as well as stationary and axisymmetric  fluid and scalar field configurations.
Keeping only first-order terms in the angular velocity $\Omega=u^{\phi}/u^{t}$, the spacetime metric
can be written in the form \cite{Hartle1967}

\begin{eqnarray}
ds^2_{*}= - e^{2\phi(r)}dt^2 + e^{2\Lambda(r)}dr^2 + r^2(d\theta^2 +
\sin^2\theta d\vartheta^2 ) - 2\omega(r,\theta)r^2 sin^2\theta  d\vartheta dt.
\end{eqnarray}

The last term in the expression for the metric reflects the influence of the rotation in linear order in $\Omega$, i.e. $\omega \sim {\cal O}(\Omega)$.
The effect of rotation on the other metric functions and the scalar field is of order ${\cal O}(\Omega^2)$. The influence of the rotation on the fluid
energy density and pressure is also of order ${\cal O}(\Omega^2)$. For the fluid four-velocity  $u^{\mu}$, up to linear terms in $\Omega$, one finds
$u=u^{t}(1, 0, 0, \Omega)$, where $u^{t}=e^{-\Phi(r)}$.

Taking into account all the symmetries and conditions imposed above, the dimensionally reduced Einstein frame field equations containing at most terms linear in $\Omega$,  are the following

\begin{eqnarray}
&&\frac{1}{r^2}\frac{d}{dr}\left[r(1- e^{-2\Lambda})\right]= 8\pi G
A^4(\varphi) \rho + e^{-2\Lambda}\left(\frac{d\varphi}{dr}\right)^2
+ \frac{1}{2} V(\varphi), \label{eq:FieldEq1} \\
&&\frac{2}{r}e^{-2\Lambda} \frac{d\phi}{dr} - \frac{1}{r^2}(1-
e^{-2\Lambda})= 8\pi G A^4(\varphi) p +
e^{-2\Lambda}\left(\frac{d\varphi}{dr}\right)^2 - \frac{1}{2}
V(\varphi),\label{eq:FieldEq2}\\
&&\frac{d^2\varphi}{dr^2} + \left(\frac{d\phi}{dr} -
\frac{d\Lambda}{dr} + \frac{2}{r} \right)\frac{d\varphi}{dr}= 4\pi G
\alpha(\varphi)A^4(\varphi)(\rho-3p)e^{2\Lambda} + \frac{1}{4}
\frac{dV(\varphi)}{d\varphi} e^{2\Lambda}, \label{eq:FieldEq3}\\
&&\frac{dp}{dr}= - (\rho + p) \left(\frac{d\phi}{dr} +
\alpha(\varphi)\frac{d\varphi}{dr} \right), \label{eq:FieldEq4} \\
&&\frac{e^{\Phi-\Lambda}}{r^4} \partial_{r}\left[e^{-(\Phi + \Lambda)} r^4 \partial_{r}{\bar\omega} \right]  + \frac{1}{r^2\sin^3\theta} \partial_{\theta}\left[\sin^3\theta\partial_{\theta}\bar\omega \right]= 16\pi GA^4(\varphi)(\rho + p)\bar\omega ,
\end{eqnarray}
where we have defined

\begin{eqnarray}
\alpha(\varphi)= \frac{d\ln A(\varphi)}{d\varphi}=-\frac{1}{\sqrt{3}} \;\;\; {\rm and}\;\;\;  \bar\omega = \Omega - \omega.
\end{eqnarray}
Here $p$ and $\rho$ are the pressure and energy density in the Einstein frame and they are connected to the Jordan frame quantities $p_*$ and $\rho_*$ via $\rho_{*}=A^{4}(\varphi)\rho$ and $p_{*}=A^{4}(\varphi) p$ respectively.

The above system of equations,  supplemented with the equation of state for the star matter and appropriate boundary conditions, describes the interior  and the exterior of
the neutron star. Evidently in the exterior of the neutron star we have to set $\rho=p=0$.

What is important for this system of equations is the fact that the equation for $\bar\omega$ is separated from the other equations which form an independent subsystem.
This subsystem is just the system of reduced field equations for the static and spherically symmetric case. The natural boundary conditions
at the center of the star are $\rho(0)=\rho_{c}, \Lambda(0)=0,$  while at infinity we have $\lim_{r\to \infty}\phi(r)=0, \lim_{r\to \infty}\varphi
(r)=0$ as required by the asymptotic flatness \cite{Yazadjiev2014}. The coordinate radius $r_S$ of the star is determined by the
condition $p(r_S)=0$ while  the physical radius of the star as measured in the physical (Jordan) frame is given by $R_{S}= A[\varphi(r_S)] r_S$.

The equation for $\bar \omega$ is in fact an elliptical partial differential equation on a spherically symmetric background. This fact and the asymptotic behaviour of $\bar \omega$ at spacial infinity
allow us to considerably simplify this equation. Expanding $\bar \omega$ in the form \cite{Hartle1967}

\begin{eqnarray}
\bar\omega= \sum^{\infty}_{l=1}{\bar \omega}_{l}(r) \left(- \frac{1}{\sin\theta}\frac{dP_{l}}{d\theta} \right),
\end{eqnarray}
where $P_{l}$ are Legendre polynomials and substituting into the equation for $\bar \omega$ we find

\begin{eqnarray}\label{OL}
\frac{e^{\Phi-\Lambda}}{r^4} \frac{d}{dr}\left[e^{-(\Phi+ \Lambda)}r^4 \frac{d{\bar\omega}_{l}(r)}{dr} \right] - \frac{l(l+1)-2}{r^2} {\bar\omega}_{l}(r)=
16\pi G A^4(\varphi)(\rho + p){\bar\omega}_{l}(r).
\end{eqnarray}

In asymptotically flat spacetimes, the asymptotic of the exterior solution of  (\ref{OL}) is ${\bar \omega}_{l} \to {\rm const}_1\, r^{-l-2} + {\rm const}_2\, r^{l-1}$. Taking into account that
$\omega \to 2J/r^3$ (or equivalently $\bar\omega \to \Omega - 2J/r^3 $ ) for $r\to \infty$ with $J$ being the angular momentum of the star and comparing it with de above asymptotic for $\bar \omega$, we conclude that $l=1$, i.e.
${\bar \omega}_{l}=0$ for $l\ge 2$. In other words, $\bar\omega$ is a function of $r$ only. Therefore the equation for the $\bar \omega$  is

\begin{eqnarray}\label{OR}
\frac{e^{\Phi-\Lambda}}{r^4} \frac{d}{dr}\left[e^{-(\Phi+ \Lambda)}r^4 \frac{d{\bar\omega}(r)}{dr} \right] =
16\pi G A^4(\varphi)(\rho + p){\bar\omega}(r).
\end{eqnarray}

The natural boundary conditions for this equation are

\begin{eqnarray}
\frac{d{\bar\omega}}{dr}(0)= 0 \;\;\; {\rm and}\;\;\; \lim_{r\to \infty}{\bar\omega}=\Omega .
\end{eqnarray}
The first condition ensures the regularity of $\bar\omega$ at the center of the star.

One of the most important quantities we consider in the present paper is the inertial moment $I$ of the  compact star. It is defined in the standard way

\begin{eqnarray}
I=\frac{J}{\Omega}.
\end{eqnarray}
Using  equation (\ref{OR}) for $\bar \omega$  and the asymptotic form of $\bar \omega$ one can find another expression for the inertial moment, namely

\begin{eqnarray}\label{eq:I_integral}
I= \frac{8\pi G}{3} \int_{0}^{r_S}A^4(\varphi)(\rho + p)e^{\Lambda - \Phi} r^4 \left(\frac{\bar\omega}{\Omega}\right) dr .
\end{eqnarray}

In the next section where we present our numerical results we shall use the dimensionless parameter $a\to a/R^2_{0}$  and the dimensionless inertial moment
$I\to I/M_{\odot}R^2_{0} $ where $M_{\odot}$ is the solar mass and $R_{0}$ is one half of the solar gravitational radius   $R_{0}=1.47664 \,km$ (i.e. the solar mass in geometrical units).

\section{Results}
The last supplement in solving the field equations \eqref{eq:FieldEq1}--\eqref{eq:FieldEq4},\eqref{OR} is to specify the EOS. We will be working with two classes of EOSs -- realistic hadronic EOS and quark matter EOS. Let us note that these are the first self-consistent and non-perturbative with respect to $a$  results of strange stars in  $f(R)$ theories of gravity, even in the static case. The hadronic EOSs are
Sly4 \cite{Douchin2001} and APR4 \cite{AkmalPR}, which both reach the two solar mass barrier and are in agreement with the current estimates of the neutron star radii \cite{Lattimer12,Steiner2010,Ozel2013,Antoniadis13,Demorest10}. The strange matter EOS is taken to be of the form

\begin{equation} \label{EQ:quark_EOS}
p=b(\rho-\rho_0) ,
\end{equation}
where $b$ and $\rho_0$ are parameters obtained from fitting different quark star EOS. In this paper we will work with $b=1/3$ and $\rho_0=4.2785 \times 10^{14} {\rm g/cm^3}$, which corresponds to the SQSB60 equation of state given in \cite{Gondek-Rosinska2008}. This strange star EOS is just a little bit below the two solar mass barrier, but we consider it as a representative example and we expect that other strange matter EOS will lead to qualilatively similar results.

The presence of a nontrivial potential of the scalar field makes the system of differential equations stiff, as we commented in \cite{Yazadjiev2014}. In addition, as it is evident from equation (\ref{EQ:quark_EOS}), the surface density for quark stars is different from zero (more precisely it is equal to $\rho_0$), which naturally causes a discontinuity of the density function at the surface and makes the calculations nontrivial. Hence one should introduce certain modifications in the applied shooting procedure and keep a close look at the numerical calculations.

\begin{figure}[]
\centering
\includegraphics[width=1\textwidth]{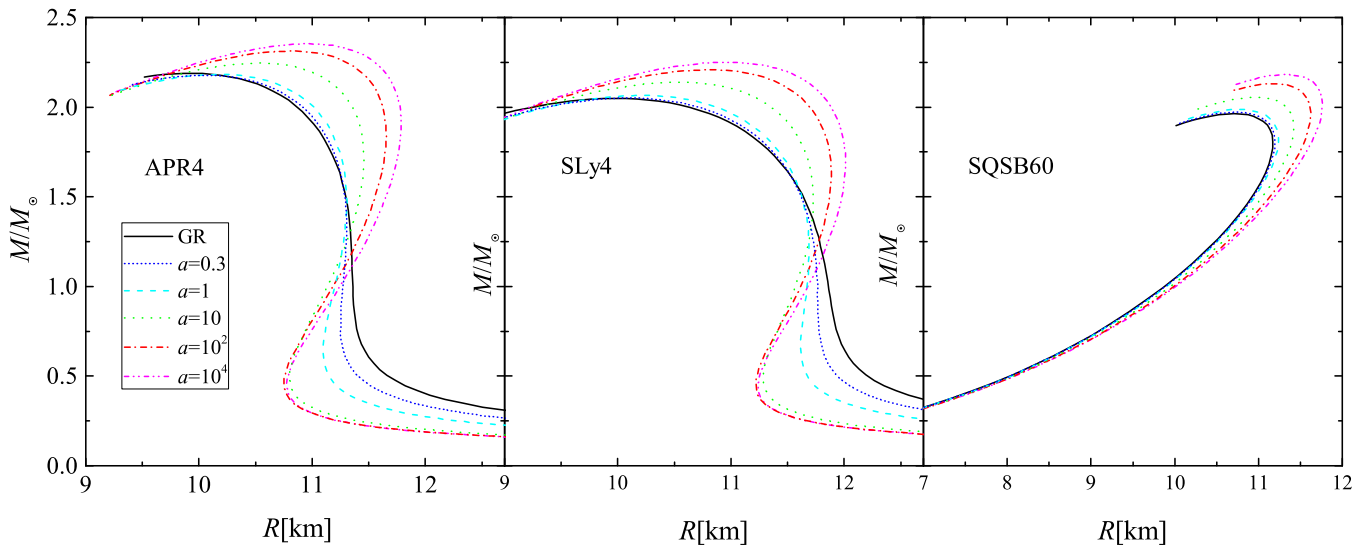}
\caption{The mass of radius diagram for EOS APR4 (left panel),
SLy4 (middle panel) and the strange star EOS (right panel). Different styles and colors of the curves
correspond to different values of the parameter  $a$.}
\label{Fig:M(R)_All}
\end{figure}

The mass as a function of the stellar radius is shown in Fig. \ref{Fig:M(R)_All} for the two hadronic EOSs and the strange star EOS. As we commented, the masses and radii in the slow rotation approximation we are considering are the same as in the static case, because the changes in $M$ and $R$ are of order $\Omega^2$. A detailed study of the mass-radius diagram, including a comparison with the available observational constraints, was made in \cite{Yazadjiev2014} in the case of hadronic EOS. The behavior of the strange stars is somehow different -- for masses below roughly $1M_\odot$, the deviations from general relativity are almost negligible, and strong deviations are observed only close to the maximum mass. Also the radius of strange stars in $f(R)$ gravity is always larger compared to the general relativistic case, while for neutron stars it is larger for massive models and smaller for less massive ones. The increase of the maximum strange mass with respect to the Einstein's theory of gravity is around $10\%$, similar to the neutron stars case.

\begin{figure}[]
\centering
\includegraphics[width=1\textwidth]{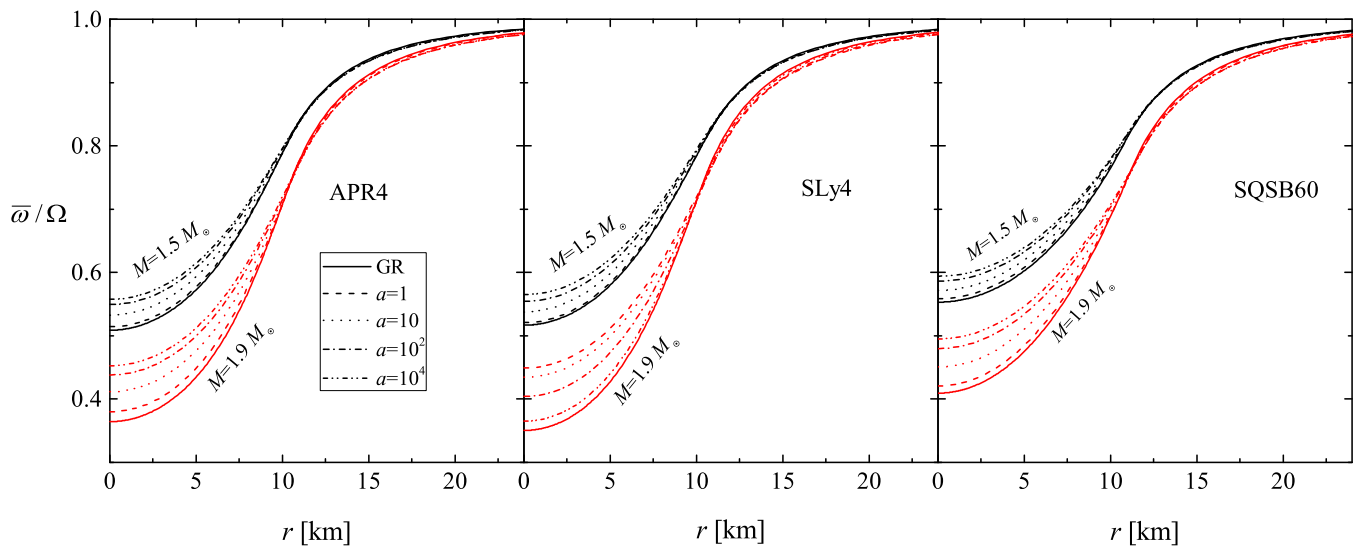}
\caption{The normalized metric function ${\bar\omega}/\Omega$ as a function of the radial coordinate for EOS APR4 (left panel), EOS SLy4 (middle panel) and the strange star EOS (right panel). Models with two different values of the mass, $M=1.5M_\odot$ and $M=1.9M_\odot$, are shown with different line colors. Different patterns of the lines correspond to different values of the parameter $a$.}
\label{Fig:OM(M)_All}
\end{figure}

We proceed now to the effects of rotation. In Fig. \ref{Fig:OM(M)_All} we plot  the normalized metric function ${\bar\omega}/\Omega$ as a function of the radial coordinate for different values of the parameter $a$ and for models with two different fixed values of the mass -- $M=1.5M_\odot$ and $M=1.9M_\odot$. The deviations from pure general relativity are stronger towards the center of the star, where the main contribution  of the scalar field is. The differences decrease when we tend towards the limit at asymptotic infinity ${\bar\omega}/\Omega\rightarrow 1$. Also larger masses normally lead to larger differences compared to GR. As should be expected, decreasing of the parameter $a$ leads to a decrease of the central value for $\bar \omega$ and in the case $a\rightarrow0$ they coincide with GR (see \cite{Yazadjiev2014} for more details). The maximum reached deviation (for $a\rightarrow \infty$) is very close to the $a=10^4$ case on the graph. For very small parameters $a$ (typically below $a<0.5$) the central value of the function ${\bar\omega}/\Omega$ might fall slightly below the general relativistic case, but these solutions differ only marginally from GR and they are not shown on the graph.

We can compare our results to the case of scalar-tensor theories and tensor-vector-scalar theories of gravity \cite{Sotani2010,Sotani2012}. In the particular scalar-tensor theory considered in \cite{Sotani2010}, the qualitative behavior of $\bar \omega$ is the same, but if we restrict ourselves to values of the coupling parameters allowed by the observations of binary neutron stars \cite{Freire2012,Antoniadis13}, then the results in scalar-tensor theories will be almost indistinguishable from general relativity. The tensor-vector-scalar theories give a different qualitative behavior of the function ${\bar\omega}/\Omega$ -- its central value decreases with respect to GR for the whole range of studied parameter, but these deviations are small and within the range of the  EOS uncertainties. Only the $f(R)$ theories can give a significant changes of the function $\bar \omega$ for large values of $a$ -- the differences in the central values of ${\bar\omega}/\Omega$ reach approximately 30\%, that is much larger than the values reported in \cite{Sotani2010,Sotani2012}.

\begin{figure}[]
\centering
\includegraphics[width=1\textwidth]{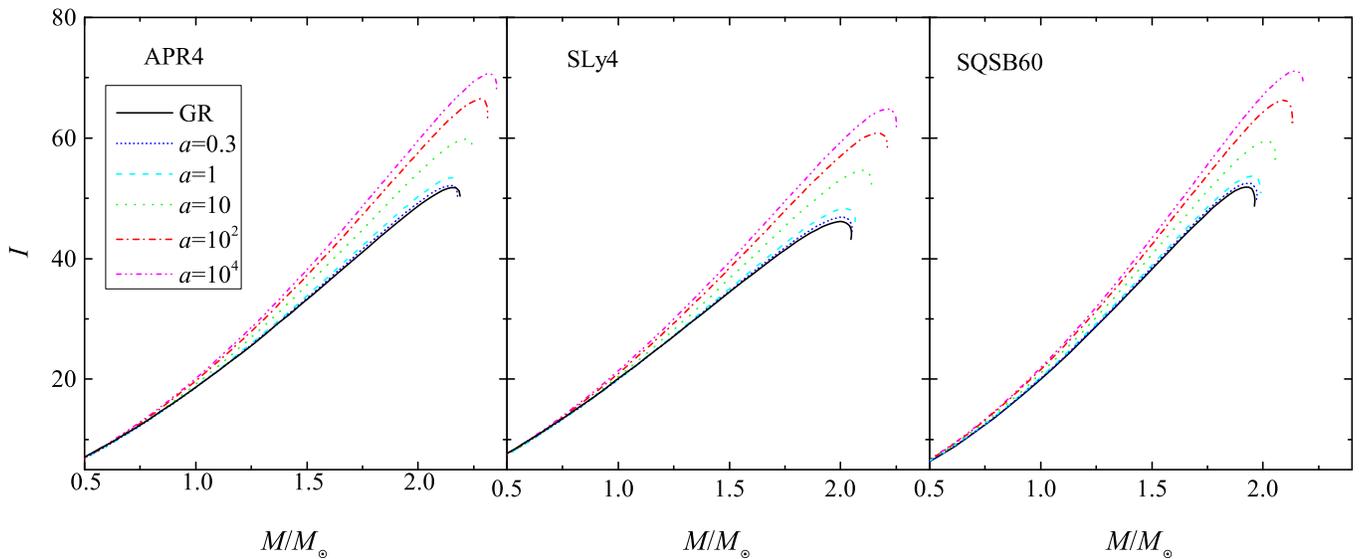}
\caption{Moment of inertia as a function of mass for realistic equations of state -- EOS APR4 in
 the left panel, EOS SLy4 in the middle panel and the strange star EOS in the right panel. Different values of the
parameter $a$ are plotted with different line styles and colors.}
\label{Fig:I(M)_All}
\end{figure}

A global characteristic of neutron stars, that can be obtained via the slow rotation approximation, is the moment of inertial given by equation (\ref{eq:I_integral}). In Fig. \ref{Fig:I(M)_All} we plot the neutron star moment of inertia as a function of mass for different values of the parameter $a$. In these graphs we see the well know behavior -- when decreasing the parameter $a$, the values of the moment of inertia get closer to the GR case and in the limiting case of $a\rightarrow 0$ they coincide. The maximum value of the moment of inertia for stable neutron star models in $R^2$ theories increases with up to 40\% with respect to the GR case for a fixed EOS, as Fig. \ref{Fig:I(M)_All} shows. This is much larger compared to the mass-radius relations, where the increase of the maximum mass for $R^2$ gravity does not exceed roughly 10\%. One can also notice that the increase of the moment of inertia is much stronger than the EOS uncertainty, even if we include the strange star equation of state. The observations of the moment of inertia in double neutron stars, on the other hand, are expected to reach an uncertainty of roughly 10\% in the next few years \cite{Lattimer2005,Kramer2009}. This would help us to set very tight constraints on the parameter $a$ in the $R^2$ gravity we are considering, and similar constraints can be derived also for other classes of $f(R)$ gravity. Currently the tightest observational constraint on $a$ comes from Gravity Probe B experiment, namely  $a\lesssim 2.3 \times 10^5$ (or $a\lesssim 5\times 10^{11} m^2$ in physical units) \cite{Naf2010}, which means that the future observations of the neutron star moment of inertia could improve this estimate by orders of magnitude.

\begin{figure}[]
\centering
\includegraphics[width=1\textwidth]{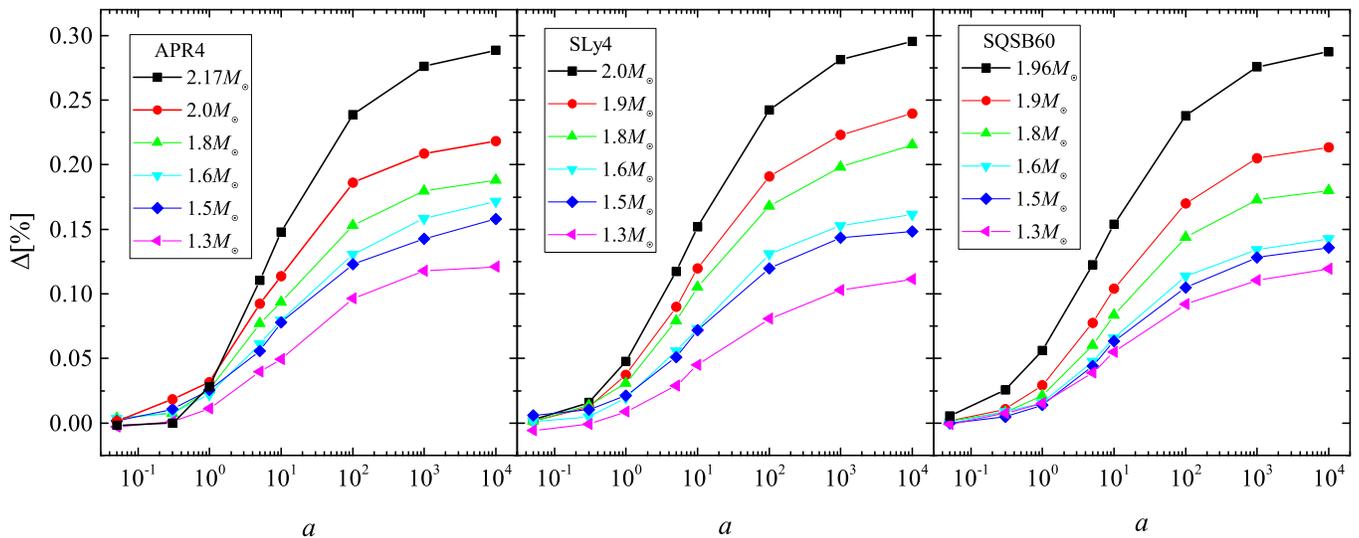}
\caption{The relative deviation $\Delta$ as a function of the parameter $a$ for several fixed values of the neutron star mass. The results for EOS APR4 are give in the left panel, for EOS SLy4 -- in the middle panel and for strange stars in the right panel.}
\label{Fig:Delta_All}
\end{figure}

In order to quantify better the changes in the moment of inertia induced by $f(R)$ theories, we can define a relative difference $\Delta$ between the moment of inertia of models with equal mass in GR and $R^2$ gravity:
\begin{equation} \label{EQ:relativ_dev}
\Delta = \left.\frac{I_{f(R)}-I_{GR}}{I_{GR}}\right|_{M={\rm const}}.
\end{equation}
Such differences were also examined in the case of scalar-tensor theories and TeVeS in \cite{Sotani2010,Sotani2012}.
The value of $\Delta$ as a function of the parameter $a$ for models with different fixed masses, are shown on Fig. \ref{Fig:Delta_All}. As we can see, for small values of the parameter $a$, the deviation
is close to zero and it can even reach negative values. But with the increase of $a$  the deviation grows, reaching up to about $30\%$ for $a=10^4$. The masses plotted on Fig. \ref{Fig:Delta_All} reach the maximum mass of the GR solutions for the corresponding EOS \footnote{As we commented, neutron stars in $f(R)$ gravity can reach larger maximum mass and thus the moment of inertia can increase further.}. It is evident that larger deviations from GR are accomplished for larger masses. Thus in order to achieve differences beyond the expected observational uncertainty, one has to consider masses above approximately $1.5M_\odot$.

\section{Conclusions}
In the present paper we considered slowly rotating neutron stars in $R^2$ gravity, keeping only terms of first order in the angular velocity $\Omega$. The calculations are made non-perturbatively with respect to the parameter $a$ and self-consistently, and they are a natural extension of the static case considered in \cite{Yazadjiev2014}. For this purpose we derived the necessary field equations and solve them numerically.

In our studies we consider three equations of state -- two hadronic EOS which are both in agreement with the current observational constraints, and a strange star EOS. The mass and the radius of neutron stars in slow rotation approximation remain unchanged with respect to the static case, but it is interesting to comment on the qualitative difference between strange stars and neutron stars, as strange stars in $f(R)$ gravity were not considered until now within the non-perturbative approach with respect to $a$. The most important difference is that below approximately $1M_\odot$ the mass-radius relation for strange stars in $R^2$ theories differs only marginally from the GR case. This is qualitatively different from neutron stars, where the radius for small masses can decrease considerably with respect to GR. For both classes of EOS, the increase of the maximum masses in $f(R)$ theories can reach up to approximately $10\%$ in the limit $a\rightarrow \infty$ (more details on the hadronic EOS are given in \cite{Yazadjiev2014}).

One of the important quantities, the slow rotation approximation can provide, is the moment of inertia. The maximum value of the moment of inertia for stable neutron star models, that can be reached in $R^2$ gravity for very large values of $a$, more precisely for $a\gtrsim 10^4$, is approximately $40\%$ larger than the general relativistic case. This is a significant deviation compared to the change in the maximum mass and it is beyond the EOS uncertainly (even if we include the strange matter EOS). If we consider sequences of models with fixed masses, but different values of $a$, the difference between the moment of inertia for very large $a$ ( $a\gtrsim 10^4$)  and for $a\rightarrow 0$ (GR) is up to roughly $30\%$. In general larger masses produce larger deviations. If we assume that the future observations of double pulsars will be able to reach 10\% accuracy \cite{Lattimer2005,Kramer2009}, the deviations due to $R^2$ gravity will exceed the observational uncertainties for masses above roughly $1.5M_\odot$. Therefore the present results can be used to set tight constrains on the parameter $a$ in $R^2$ gravity, and this constraints will be several orders of magnitude stronger that the Gravity Probe B results. Similar investigations and conclusions can be made for other classes of $f(R)$ theories and such a study is underway. It is important to note that the deviations observed in the present paper exceed significantly the estimates in other alternative theories of gravity \cite{Sotani2010,Sotani2012}.

The radial profile of the metric function $\bar \omega$ also changes considerably. The deviations from GR are stronger in the central region of the star (where the scalar field is stronger) and for larger neutron star masses. The difference in the central value reaches as much as 30\%  for very large $a\gtrsim 10^4$. This is considerably higher than the results in other alternative theories of gravity \cite{Sotani2010,Sotani2012} and the possible astrophysical manifestations should be studied in the future.

From  all the presented results one can conclude that the $R^2$ gravity leads to very distinct neutron star characteristics even in the slowly rotating case. The rapid rotation on the other hand is known to amplify the differences from GR in the case of scalar-tensor theories \cite{Doneva2013}. As the $f(R)$ theories are mathematically equivalent to scalar-tensor theories with a potential of the scalar field, we believe that similar conclusion can be made also for rapidly rotating neutron stars in $f(R)$ theories. Even more, we expect that the deviations can be much stronger than the scalar-tensor theory case, since the changes in the static equilibrium properties are also more pronounced \cite{Damour1993,Yazadjiev2014}.
\section*{Acknowledgements}

DD would like to thank the Alexander von Humboldt Foundation for a stipend. KK,  SY and KS would like to thank the
Research Group Linkage Programme of the Alexander von Humboldt Foundation for the support. The support by
the Bulgarian National Science Fund under Grant DMU-03/6, by the Sofia University Research Fund under Grant
63/2014 and by the German Science Foundation (DFG) via SFB/TR7 is gratefully acknowledged. Partial support
comes from "New-CompStar", COST Action MP1304.

%%%%%%%%%%%%%%%%%%%%%%%%%%%%%%%%%%%%%%%%%%%%%%%%%%%%%%%%%%%%%%%%%%%%%%%%%%%%%%%

\bibliography{references}

\end{document}